\documentclass{IEEEcsmag}

\usepackage[colorlinks,urlcolor=blue,linkcolor=blue,citecolor=blue]{hyperref}
\usepackage{xcolor}
\usepackage{upmath}
\usepackage{listings}
\usepackage{balance}
\usepackage{booktabs}

\jvol{XX}
\jnum{XX}
\paper{XX}
\jmonth{October}
\jname{XX}
\pubyear{2020}

\begin{document}

%\sptitle{Department: IEEEcsmag}
%\editor{Editor: Name, xxxx@email}

\title{Reproducing GW150914: \\ the first observation of gravitational waves from a binary black hole merger}

\author{Duncan A. Brown}
\affil{Syracuse University}

\author{Karan Vahi}
\affil{University of Southern California}

\author{Michela Taufer}
\affil{University of Tennessee Knoxville}

\author{Von Welch}
\affil{Indiana University}

\author{Ewa Deelman}
\affil{University of Southern California}

%\markboth{Department Head}{Paper title}

\begin{abstract}
In 2016, LIGO and Virgo announced the first observation of gravitational waves from a binary black hole merger, known as GW150914. To establish the confidence of this detection, large-scale scientific workflows were used to measure the event's statistical significance. They used code written by the LIGO/Virgo and were executed on the LIGO Data Grid. The codes are publicly available, but there has not yet been an attempt to directly reproduce the results, although several analyses have replicated the analysis, confirming the detection. We attempt to reproduce the result presented in the GW150914 discovery paper using publicly available code on the Open Science Grid. We show that we can reproduce the main result but we cannot exactly reproduce the LIGO analysis as the original data set used is not public. We discuss the challenges we encountered and make recommendations for scientists who wish to make their work reproducible.
\end{abstract}

\maketitle

%% Sections
\chapterinitial{For the scientific community} 
\label{sec:introduction}
to build on previous results, it must trust that these results are not accidental or transient, but rather that they can be reproduced to an acceptably high degree of similarity by subsequent analyses. This notion of reproducibility is magnified both in importance and challenges in the context of computational science workflows~\cite{Stodden:2016}. An increasingly large fraction of scientific results depend on computational elements, which in turn creates reproducibility challenges associated with the implementation of these computational elements. 
Being able to reason about the validity of published scientific results and re-use them in derivative works becomes an extremely challenging task. Publishers have made great strides in including relevant artifacts along with the manuscripts. However, data, methods, and results are still hard to find and harder still to reproduce (re-creating the results from the original author's data and code), to replicate (arriving at the same conclusion from a study using new data or different methods), and to re-use in derivative works (using code or data from a previous study in a new analysis)~\cite{Heroux:2018}. 

Our work focuses on reproducing the computational analysis used to establish the significance of the first detection of gravitational waves created colliding  binary black holes and observed by the Advanced Laser Interferometer Gravitational-wave Observatory (LIGO)~\cite{Abbott:2016blz}. 
As part of its commitment to Open Data, LIGO made the data and scientific codes from its first observing run available to the  scientific community.
Previous analyses have \emph{replicated} the results of the GW150914 discovery~\cite{Venumadhav:2019tad,Nitz:2018imz}. In these analyses, the data from LIGO's first observing run was re-analyzed either by independent teams of scientists with different codes, with different data, or by using different workflows to those used in the original GW150914 discovery. 

In a previous work, we have performed a \emph{post-hoc} comparison of these results using the published papers and the PRIMAD reproducibility formalism~\cite{Chapp:2019}. Here, we attempt to reproduce \emph{ab-initio} the original LIGO analysis used in the GW150914 discovery paper using public information. Specifically, we attempt to reproduce the results of the PyCBC search for gravitational waves~\cite{Usman:2015kfa,TheLIGOScientific:2016qqj} shown in Figure 4 of Abbott {\em et al.}~\cite{Abbott:2016blz}. 

Our effort is not completely separate from the original analysis, as one co-author of this paper was a member of the team involved in running the original LIGO analysis. However, our aim was to automate the production of the result in a way that other co-authors of this paper who were not members of the LIGO or Virgo collaborations, as well as other scientists, could reproduce the result. 

The original analysis workflows 
were executed on the LIGO Data Grid, a collection of computational resources that are not available to the wider community. Since non-LIGO scientists do not have access to these systems, we execute the analysis on the Open Science Grid (OSG)~\cite{Bockelman:2015} and rely on a cyberinfrastructure software stack that has latest stable releases of key software packages such as HTCondor~\cite{Bockelman:2015}, Pegasus~\cite{Deelman:2019}, and the CERN Virtual Machine Filesystem (CVMFS)~\cite{Weitzel:2017}. We have created a script that automates the setup and deployment of the LIGO workflows on a typical local compute cluster 
and from there Pegasus manages their execution on OSG.

\begin{figure*}[!t]
\begin{centering}
\includegraphics[width=0.45\textwidth]{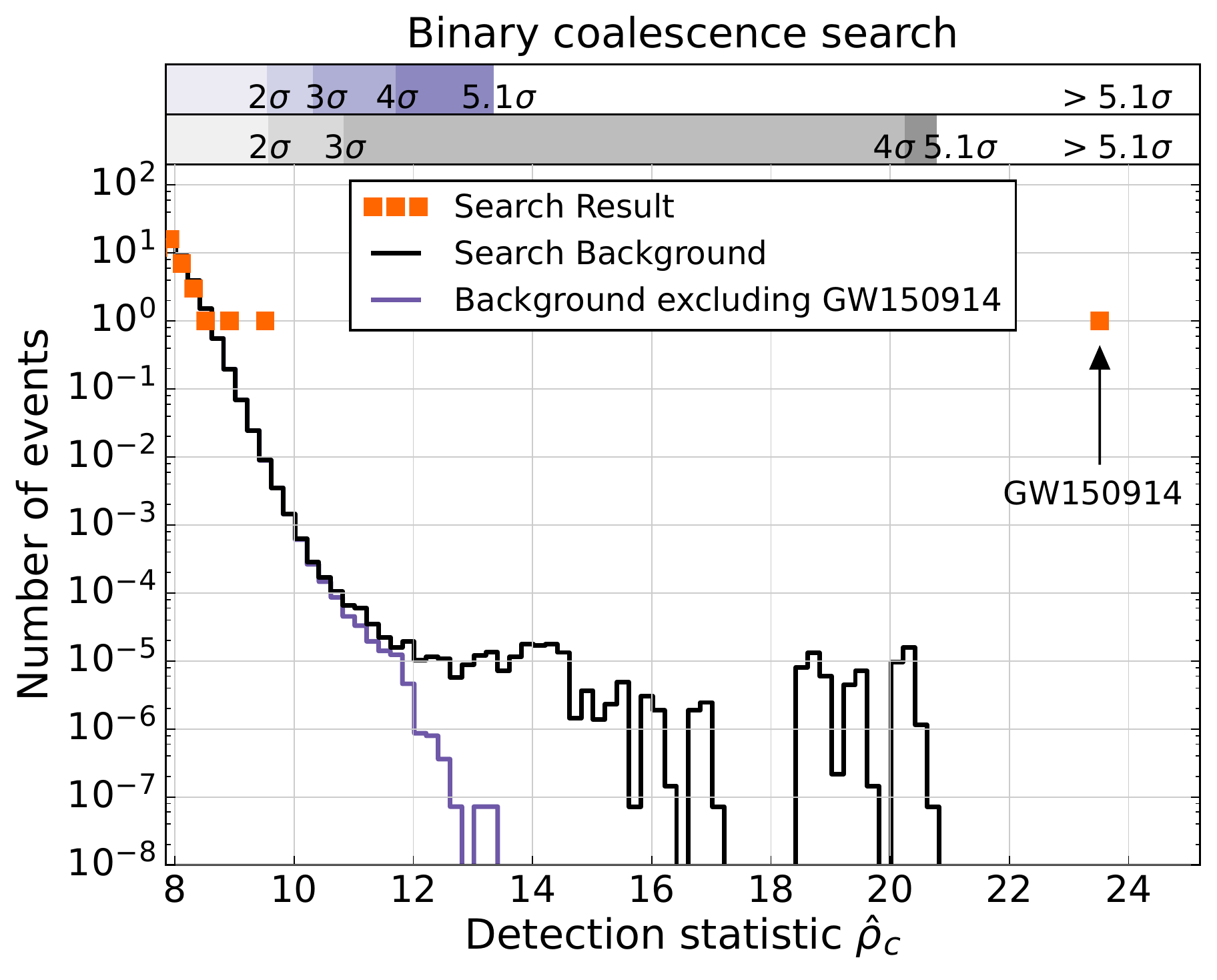}
\includegraphics[width=0.45\textwidth]{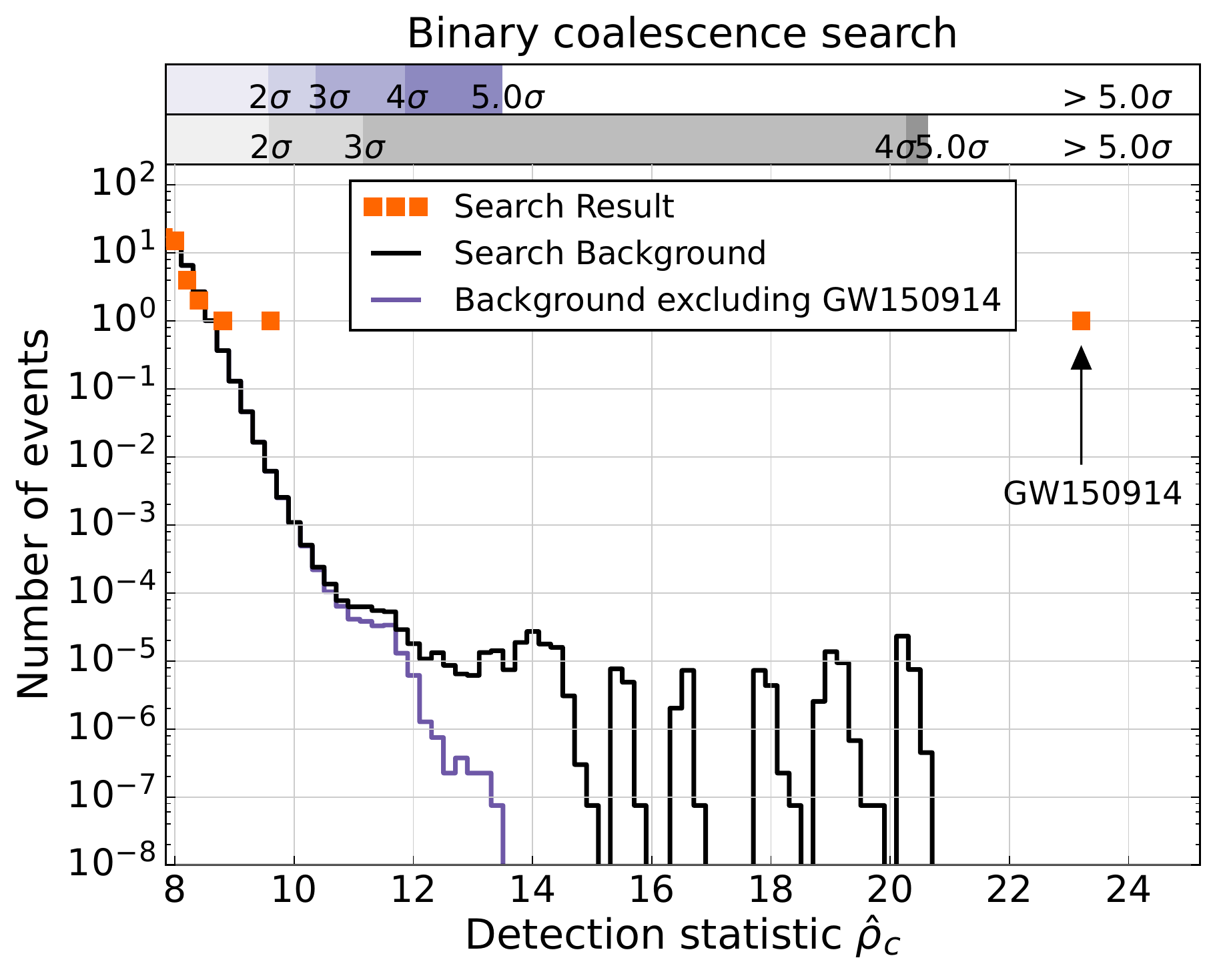}
\end{centering}
\caption{
Results from the binary coalescence search presented in the GW150914 discovery paper from \cite{Abbott:2016blz} with permission (left) and our attempt to reproduce these results (right). These histograms show the number of candidate events (orange markers) and the mean number of background events (black lines) as a function of the search detection statistic and with a bin width of 0.2. The scales on the top give the significance of an event in Gaussian standard deviations based on the corresponding noise background. We were able to reproduce the search result for GW150914, but we were unable to exactly reproduce the search background. The differences between the two figures is likely due to differences in the gravitational-wave stain data used, as described in the text. 
\label{f:gw150914-fig4b}
}
\end{figure*}

Our main goal was to reproduce the results of the PyCBC search shown in Figure 4 of Abbott {\em et al.}~\cite{Abbott:2016blz}, shown on the left side of our {\bf Figure \ref{f:gw150914-fig4b}}, since this is the result used to make the statement that the signal is detected with a ``significance greater than $5.1~\sigma$ in the abstract of the paper. Our reproduction of this plot, shown on the right-hand side of Figure \ref{f:gw150914-fig4b} shows that we can reproduce the search result, but there are small, noticeable differences in the search background (explained later in the paper). Based on the LIGO documentation, we believe that these differences are because the data used in the original analysis were different from that released by the Gravitational Wave Open Science Center (GWOSC)~\cite{Vallisneri:2014vxa} and used in our analysis. Unfortunately, the original data set is not public and so we are unable to confirm this hypothesis. However, we consider our ability to re-run a scientific workflow last executed in 2015 and largely reproduce the results to be a successful demonstration of reproducibility. 

This article is structured as follows. First, we provide background on the first gravitational-wave discovery. We then describe our recent efforts on ab-inito analysis to reproduce the GW150914 result, followed by challenges we encountered. In the results section, we explain any difference observed in our reproduction of result published by LIGO. We  perform an analysis of the workflow run-time provenance data and the compute resources required to execute the workflow.  We conclude with recommendations for others who wish to reproduce the GW150914 result.

\section{THE DISCOVERY OF GW150914}
\label{sec:discovery}

Gravitational-wave astronomy is an interesting case study for robust science because it has three main science phases: low-latency data analysis, offline analysis, and public and educational dissemination of results. The low-latency analysis processes instrumental data in near real time to identify astrophysical signals. Alerts are disseminated to the community to identify electromagnetic or neutrino counterparts to the gravitational-wave signal. Offline analyses validate the low-latency detections, identify signals missed in low-latency, and provide determination of source properties.  When a detection is published, the data is released to the scientific community and the public. Since the analysis codes are also released, it should be possible for people outside the LIGO Scientific Collaboration and Virgo to reproduce the published results.  

Our attempt to reproduce the  first detection of gravitational waves from binary black holes, known as GW150914, starts from the data released by the GWOSC~\cite{Vallisneri:2014vxa}. GW150914 was first detected by a low-latency search for gravitational-wave bursts that identifies interesting candidates but does not provide the final statistical significance of detected events. To establish the significance of events, data from the LIGO detectors is subsequently analyzed by scientific workflows that use longer stretches of data to provide a measure of the noise background in the detectors and use this to measure the significance of candidate events. Results from two offline analyses were presented in the GW150914 discovery paper: one that used a search technique that did not make assumptions about the shape of the gravitational waveform~\cite{Abbott:2016blz} and one using matched filtering (comparing the data to a known waveform) to search for the signals from merging black holes~\cite{Usman:2015kfa}, known as PyCBC. Here, we focus on reproducing the results of the PyCBC binary black hole search.

The PyCBC search uses matched filtering to compare the LIGO data with a bank of template binary black hole waveforms that model the target sources. If the noise in the LIGO detectors was stationary and Gaussian, the estimation of the statistical significance of candidate events that crossed a signal-to-noise ratio threshold would be straightforward. However, the LIGO detector data contains non-Gaussian noise transients and periods of non-stationary noise. As a result, additional signal-processing techniques are applied to the data that 
%down-weight the signal-to-noise ratio of 
suppress non-Gaussian noise events. The search algorithms require that the same signal is seen in the detectors; the same waveform must be present both detectors and the signal's time of arrival must be consistent with the gravitational-wave travel time between the observatories. The map between the detection statistic (weighted signal-to-noise ratio) and the statistical significance of an event must be empirically measured by the workflow. This is done by time-shifting the data between the detectors and repeating the coincidence analysis many times. The most computationally-intensive part of the PyCBC workflow are the matched filtering and the calculation of the detection statistics. Performing the coincidence and the time-shift analysis can require a large amount of memory to process the candidate events. Once these steps are complete, the workflow produces a measurement of the statistical significance of candidates. A separate script run after the workflow completes produces a histogram that compares candidate events to the noise background.

\section{REPRODUCING THE ANALYSIS}
\label{sec:approach}

Our work is the first attempt to \emph{reproduce} the original LIGO analysis. Previous analyses, for example 1-OGC result~\cite{Nitz:2018imz}, provide an example of \textit{replication} in gravitational-wave science. In the 1-OGC analysis, a different team with different experimental setup recovered the discovery of GW150914. Here, ``different experimental setup'' means a modified data-analysis pipeline with a different configuration to that used in the original analysis. The 1-OGC result independently confirmed that GW150914 was a high significance discovery, but the event was recovered with slightly different parameters to the original discovery; these parameter differences can be explained by  differences between the different algorithms used.

Here, we provide an example of \textit{reproducibility} of the measurement of the statistical significance of GW150914 shown in Figure 4 of Abbott {\it et al.}~\cite{Abbott:2016blz}.  Reference~\cite{Abbott:2016blz} by its nature as a brief letter does not provide sufficient information to reproduce the result. Reference~\cite{TheLIGOScientific:2016qqj}  provides additional description of the analysis and the codes\footnote{\url{https://github.com/gwastro/pycbc}} and configuration files\footnote{\url{https://github.com/gwastro/pycbc-config}} are publicly released on GitHub. Although the codes and configuration are public, the LIGO/Virgo collaboration does not provide full instructions for running the workflow and reproducing the analysis. Our work provides a fully reproducible process.

Not all of the information needed to reproduce the GW150914 workflow was available in the public release accompanying the publications. The lack of a single, public repository of this knowledge is the most significant challenge for a group outside the LIGO and Virgo collaborations to reproducing the GW150914 result. However, one of the co-authors of our work was a member of the team who performed the original analysis. They were able to review their unpublished notes, which allowed us to successfully reproduce the LIGO analysis. To ensure that scientists who were not involved in the original analysis could reproduce the results, we created scripts that were run independently by another author of this paper not involved in the original analysis. These scripts were created in a peer-programming style, which started from the original scripts used to run the LIGO workflow and created the result plot. We iteratively fixed problems encountered when trying to run the analysis using information entirely in the public domain, filling in missing public information with the original analysis notes where necessary.

PyCBC is a gravitational-wave data-analysis toolkit written primarily in Python with C extensions for numerically-intensive computations. Re-running old versions of interpreted Python code can be challenging if the underlying software stack has changed since the code was originally executed. Fortunately, LIGO packaged the PyCBC codes used in the original analysis as PyInstaller bundles. These bundles package the Python code with a Python interpreter and the Python library dependencies allowing us to run the original codes without needing to recreate the entire software stack. Our final version of the workflow execution script is provided in a data release that accompanies this paper\footnote{\url{https://doi.org/10.5281/zenodo.4085984}} and the GitHub commit history\footnote{\url{https://github.com/gwastro/gw150914-fig4b/commits/1.1}} documents the iterative process of addressing the issues encountered, which we describe below.

\subsection{Software versions} Software provenance is critical to the reproducibility of scientific workflows. However, 
neither the discovery paper published in Physical Review Letters, nor the technical paper published in Physical Review D documented the exact version of the PyCBC code used to produce the analysis. The notes from the original run recorded that PyCBC v1.3.2 was used, and recorded the git commit hash of the configuration files used (which are stored in a separate GitHub repository).

\subsection{Open data} The original analysis used data and metadata that are proprietary to the LIGO Scientific Collaboration and the workflow used tools that queried proprietary servers to locate and access these data. Our script modifies the workflow to use the public data and services provided by GWOSC. For the metadata, we created wrapper codes that have the same command-line API as the proprietary codes and translate these to queries against the public data repositories. The format of the data-quality metdadata provided by GWOSC is different to that used in the original analysis. Information from the public LIGO technical note T1600011-v3 was used to determine how to use the public metadata in way that is as close as possible to the original metadata. LIGO publishes its public data using CVMFS
under the gwosc.osgstorage.org organization. Our script configures the workflow to use data from CVMFS, allowing us to rely on its distribution and caching capabilities when running jobs on the OSG. To allow the workflow generation script to find these data, we installed the LIGO Diskcache API to index the CVMFS files and the LIGO Datafind Server to resolve the workflow's metdatadata queries to file URLs for the CVMFS data. Configuration files for these tools are provided in our data release.

\subsection{Workflow format} To provide sufficient resources to run the workflow, we executed the computationally-intensive jobs on the OSG. This required a newer version of Pegasus workflow management system~\cite{Deelman:2019} than the version originally used to plan and execute the analysis. Our workflow generation script modifies the workflow written by PyCBC v1.3.2 to be compatible with Pegasus 4.9.3.

\subsection{Access to codes} 
Although all of the codes used to generate and run the analysis workflow were public, the script used to make the figure shown in Abbott {\it et al.}~\cite{Abbott:2016blz} was never released in the PyCBC software repository. Since one of the authors of this paper helped create this script, we were able to obtain the original code used. 
\section{WORKFLOW EXECUTION}
\label{sec:workflow}

After modifying the original workflow generation script to address the challenges described in the previous section, we attempted to reproduce the analysis. LIGO did not provide estimates of the runtimes or the resource requirements of the analysis tasks, so we executed the workflow on a combination of local and OSG resources. We used USC-ISI computers to manage the workflow and run the post-processing jobs and OSG resources to run the computationally-intensive jobs. Several challenges were encountered during our attempt to execute the workflow, as described below.

\subsection{Operating system and hardware mismatches} The PyCBC PyInstaller bundles are not true static executables nor are they packaged in a robust containerized environment like Singularity. The bundles require the appropriate C standard-library shared objects to be installed on the target machine and perform just-in-time compilation of bundled C code using the now-deprecated \texttt{scipy.weave} module. A standard set of OS libraries, the GNU C Compiler, and processor instructions was guaranteed for the original analysis as it was run on a single homogeneous LIGO Data Grid cluster. However, not all the OSG compute notes had the correct versions of C standard-library installed and some nodes lacked processor instructions (specifically, we encountered QEMU-emulated virtual machines that lacked the FMA4 instruction) that the PyCBC bundles required on the execute nodes. To address this, we used Pegasus and HTCondor matchmaking and fault tolerance functionalities and the ability to express requirements of the desired node characteristics to steer the PyCBC executables to compatible OSG compute nodes. 

\subsection{Non-deterministic memory use}
The amount of memory that the matched-filtering jobs required is determined by the data that they analyze. If LIGO data contains more non-Gaussian noise than average, more memory is required to compute signal-based vetoes and to store the resulting triggers. Since the noise is random, it is not possible to determine in advance how much memory is required for a given job. To address this, we configured HTCondor to automatically request more memory on each retried filtering job that failed. 

\subsection{Post-processing memory requirements} Several of the workflow's post-processing jobs require very large memory footprints (greater than 128~Gb). It was challenging to find machines with sufficient capacity for these jobs on OSG and so these jobs were executed on the local cluster at USC-ISI. This cluster is managed using HTCondor partitionable slots allowing a single job to request sufficient memory in our multi-core machines. Coordination with the cluster administrators was required to ensure that these resources were available.

\subsection{Long-term code archival} 
During the preparation of this paper, the LIGO Scientific Collaboration deleted the repository at \texttt{git.ligo.org} that stored the compiled PyInstaller PyCBC executables used in the original analysis. We had preserved a copy of the PyCBC v1.3.2 PyInstaller bundles prior to their deletion in an archive file on the IEEE DataPort server\footnote{\url{http://dx.doi.org/10.21227/c634-qh33}}. We have hosted an uncompressed version of this archive on a USC/ISI web server\footnote{\url{https://pegasus.isi.edu/ligo/eager/pycbc-software/v1.3.2/}} and configured our workflow generation script to download the bundles from the USC/ISI server. Preserving these executables will allow others to run using the bundles rather than having to recreate the complex PyCBC software stack from the public source code available on GitHub.

\section{RESULTS}
\label{sec:results}

Once the various issues described above had been addressed in our workflow-generation script \texttt{generate\_workflow.sh}\footnote{\url{https://doi.org/10.5281/zenodo.4085984}} and LIGO's PyCBC v1.3.2 PyInstaller bundles, we were able to reproduce the LIGO analysis workflow. The workflow contained almost 42,000 tasks).
We observed 28,676 task failures as the workflow ran approximately 155 days of badput (amount of computation time used on failed jobs). The majority of job failures were caused by PyInstaller bundles landing on incompatible nodes and the majority of badput was due to compute-intensive jobs being evicted due to using too much memory. Failing jobs were re-run using the retry on failure semantics in Pegasus and HTCondor that then steered these jobs to compatible nodes.
Listing 1 shows results retrieved from mining the Pegasus runtime provenance database. 
To execute the GW150914 workflow requires approximately 22 years of computing time (sum of duration of all jobs in the workflow with each job running on a single core). 

The workflow generates a web page with a number of diagnostic plots that are used by LIGO scientists to understand the detector state, the properties of the noise, and the results of the search. For the purpose of reproducing Figure~4 of Abbott {\it et al.}~\cite{Abbott:2016blz}, the primary data product is a 2.5 Gb HDF5 file that contains the triggers found in coincidence between the LIGO detectors, and the search background estimated using the time-slide method~\cite{Usman:2015kfa,TheLIGOScientific:2016qqj}. We have archived a compressed version of this HDF5 file on the IEEE DataPort server\footnote{\url{http://dx.doi.org/10.21227/c634-qh33}}.

\begin{figure}[t]
	\centering
	\includegraphics[scale=0.6]{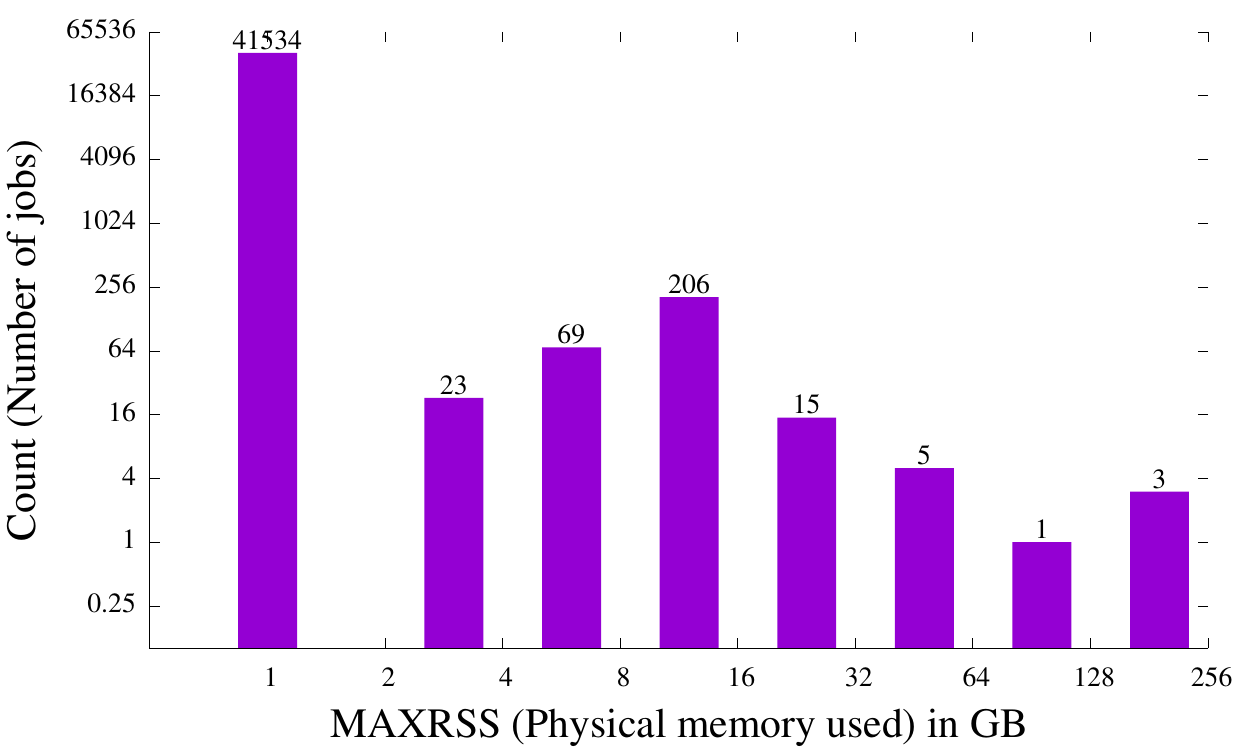}
	\caption{Frequency histogram showing maximum physical memory used by LIGO jobs as reported by \textit{pegasus-kickstart} in range of 1 to 256GB, with both X/Y axis on log scale}
    \label{fig:memory-all}
\end{figure}

\begin{figure}[!t]
	\centering
	\includegraphics[scale=0.60]{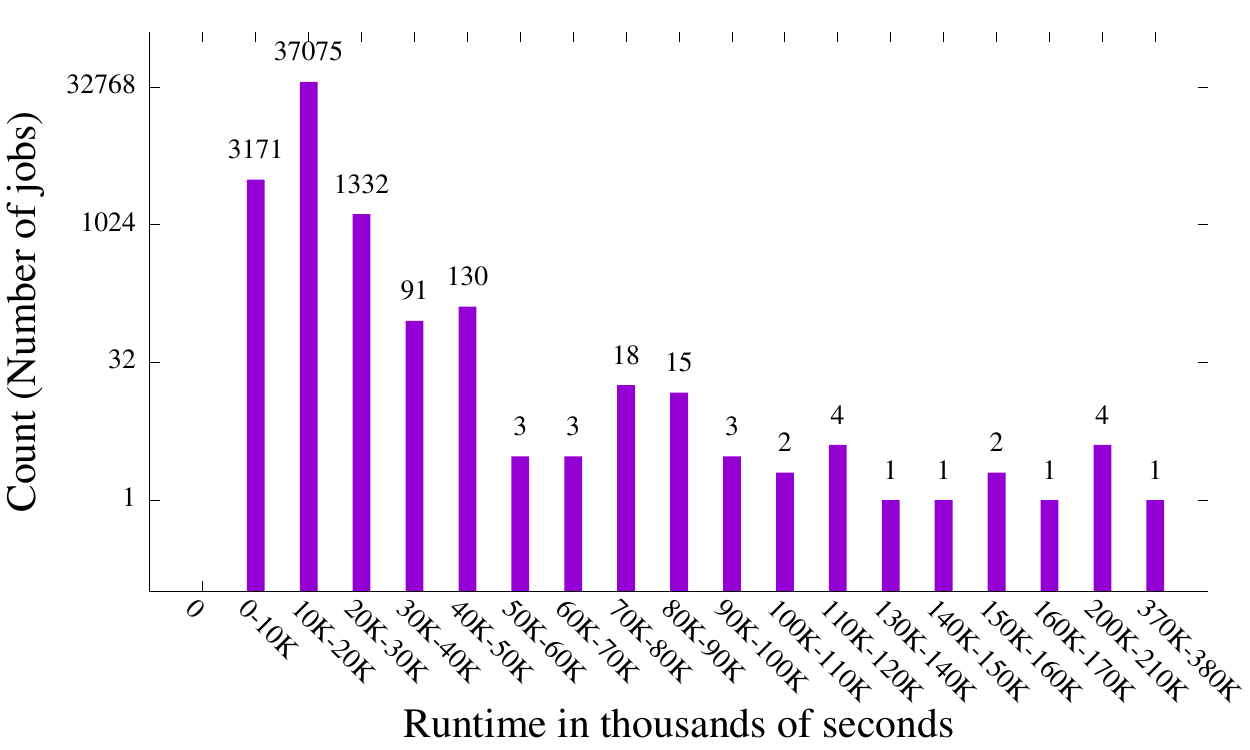}
	\caption{Frequency histogram showing runtime of LIGO jobs as reported by \textit{pegasus-kickstart} in range of 0 to 380,000 seconds}
    \label{fig:runtime}
\end{figure}

To allow future researchers to reproduce our work on their own resources, we show the distribution of physical memory used by LIGO jobs and their run-times as frequency histograms in {\bf Figure~\ref{fig:memory-all}} and {\bf Figure~\ref{fig:runtime}} respectively. In PyCBC workflows, each job type is associated with a transformation (executable). In {\bf Table~\ref{tab:top10-memory}} we show the top 10 transformations ordered by the maximum physical memory used. {\bf Table~\ref{tab:top10-runtime}} shows the top ten transformations ordered by maximum runtime in seconds.

\begin{lstlisting}[float=*, basicstyle=\small, frame=shadowbox, rulesepcolor=\color{gray}, captionpos=b, caption={Output of the pegasus-statistics tool showing runtime statistics from the OSG run}, label={lst:pegasus-statistics}]
-----------------------------------------------------------------------------
Type           Succeeded Failed  Incomplete  Total     Retries
Tasks          41856     0       0           41856     28676       
Jobs           46631     0       0           46631     28676       
Sub-Workflows  8         0       0           8         104         
-----------------------------------------------------------------------------

Workflow wall time                                       : 29 days, 0 hrs
Cumulative job wall time                                 : 22 years, 54 days
Cumulative job badput wall time                          : 155 days, 13 hrs

# Integrity Metrics
# Number of files for which checksums were compared/computed along
# with total time spent doing it. 
94713 files checksums compared with total duration of 7 hrs, 55 mins
46200 files checksums generated with total duration of 4 hrs, 9 mins

# Integrity Errors
# Total:
#       Total number of integrity errors encountered across all job 
#       executions(including retries) of a workflow.
# Failures:
#       Number of failed jobs where the last job instance had integrity errors.
Total:    A total of 54 integrity errors encountered in the workflow
Failures: 0 job failures had integrity errors
\end{lstlisting}

\begin{table*}[]
\scriptsize
\centering

\begin{tabular}{@{}llllll@{}}
\toprule
\textbf{LIGO Job Transformation}                             & \textbf{Count} & \textbf{\begin{tabular}[c]{@{}l@{}}Mean (runtime) \\ seconds\end{tabular}} & \textbf{\begin{tabular}[c]{@{}l@{}}Min (mem) \\ in MB\end{tabular}} & \textbf{\begin{tabular}[c]{@{}l@{}}Max (mem) \\ in MB\end{tabular}} & \textbf{\begin{tabular}[c]{@{}l@{}}Mean (mem) \\ in MB\end{tabular}} \\ \midrule
distribute\_background\_bins-(FDFCC)\_12H-H1L1\_ID15         & 1              & 9,673.17                                                                   & 194,898.65                                                          & 194,898.65                                                          & 194,898.65                                                           \\
statmap-(FDFCC)\_12H-H1L1\_ID16                              & 3              & 9,698.37                                                                   & 16,218.66                                                           & 189,332.06                                                          & 103,789.45                                                           \\
plot\_snrifar-(FDFCC)\_12H\_(FD\_FB)\_2-H1L1\_ID32           & 1              & 20,459.47                                                                  & 150,602.64                                                          & 150,602.64                                                          & 150,602.64                                                           \\
plot\_snrifar-(FDFCC)\_12H\_(FD\_FB)\_2\_IFAR-H1L1\_ID34     & 1              & 2,177.91                                                                   & 62,365.84                                                           & 62,365.84                                                           & 62,365.84                                                            \\
plot\_snrifar-(FDFCC)\_12H\_(FD\_FB)\_2\_CLOSED-H1L1\_ID21   & 1              & 1,654.36                                                                   & 54,009.43                                                           & 54,009.43                                                           & 54,009.43                                                            \\
plot\_snrifar-(FDFCC)\_12H\_(FD\_FB)\_0\_IFAR-H1L1\_ID24     & 1              & 1,473.62                                                                   & 40,302.30                                                           & 40,302.30                                                           & 40,302.30                                                            \\
plot\_snrifar-(FDFCC)\_12H\_(FD\_FB)\_0-H1L1\_ID22           & 1              & 1,484.56                                                                   & 40,302.11                                                           & 40,302.11                                                           & 40,302.11                                                            \\
plot\_snrifar-(FDFCC)\_12H\_(FD\_FB)\_0\_CLOSED-H1L1\_ID19   & 1              & 1,167.86                                                                   & 34,923.08                                                           & 34,923.08                                                           & 34,923.08                                                            \\
plot\_singles-MTOTAL\_EFFSPIN\_NEWSNR\_FULL\_DATA-H1\_ID47   & 1              & 1,302.16                                                                   & 32,334.82                                                           & 32,334.82                                                           & 32,334.82                                                            \\
plot\_singles-ENDTIME\_DURATION\_NEWSNR\_FULL\_DATA-H1\_ID45 & 1              & 1,168.63                                                                   & 32,334.81                                                           & 32,334.81                                                           & 32,334.81                                                            \\ \bottomrule
\end{tabular}

\caption{Top 10 LIGO job types by maximum physical memory (maxrss) used in MB, where (FDFCC) expands to FULL\_DATA\_FULL\_CUMULATIVE\_CAT and (FD\_FB) to FULL\_DATA\_FULL\_BIN}.
\label{tab:top10-memory} 
\end{table*}

\begin{table*}[]
\scriptsize
\begin{tabular}{@{}llllll@{}}
\toprule
\textbf{LIGO Job Transformation}        & \textbf{Count} & \textbf{\begin{tabular}[c]{@{}l@{}}Mean (runtime)\\  seconds\end{tabular}} & \textbf{\begin{tabular}[c]{@{}l@{}}Min (mem) \\ in MB\end{tabular}} & \textbf{\begin{tabular}[c]{@{}l@{}}Max (mem) \\ in MB\end{tabular}} & \textbf{\begin{tabular}[c]{@{}l@{}}Mean (mem) \\ in MB\end{tabular}} \\ \midrule
hdf\_trigger\_merge-FULL\_DATA-L1\_ID12 & 1              & 376,478.99                                                                 & 1,026.20                                                            & 1,026.20                                                            & 1,026.20                                                             \\
calculate\_psd-PART5-H1\_ID75           & 1              & 205,419.13                                                                 & 1,027.13                                                            & 1,027.13                                                            & 1,027.13                                                             \\
calculate\_psd-PART3-H1\_ID73           & 1              & 204,379.13                                                                 & 1,024.27                                                            & 1,024.27                                                            & 1,024.27                                                             \\
calculate\_psd-PART4-H1\_ID74           & 1              & 202,814.55                                                                 & 1,024.95                                                            & 1,024.95                                                            & 1,024.95                                                             \\
calculate\_psd-PART1-H1\_ID71           & 1              & 202,169.56                                                                 & 1,024.81                                                            & 1,024.81                                                            & 1,024.81                                                             \\
calculate\_psd-PART9-H1\_ID79           & 1              & 167,258.14                                                                 & 1,356.44                                                            & 1,356.44                                                            & 1,356.44                                                             \\
calculate\_psd-PART9-L1\_ID68           & 1              & 116,403.74                                                                 & 1,408.61                                                            & 1,408.61                                                            & 1,408.61                                                             \\
calculate\_psd-PART1-L1\_ID60           & 1              & 115,357.57                                                                 & 1,386.77                                                            & 1,386.77                                                            & 1,386.77                                                             \\
calculate\_psd-PART0-L1\_ID59           & 1              & 110,270.45                                                                 & 1,025.48                                                            & 1,025.48                                                            & 1,025.48                                                             \\
calculate\_psd-PART7-L1\_ID66           & 1              & 109,125.78                                                                 & 1,398.72                                                            & 1,398.72                                                            & 1,398.72                                                             \\
calculate\_psd-PART8-H1\_ID78           & 1              & 71,150.19                                                                  & 1,344.01                                                            & 1,344.01                                                            & 1,344.01                                                             \\ \bottomrule
\end{tabular}

\caption{Top 10 LIGO job types by max runtime in seconds}.
\label{tab:top10-runtime} 
\end{table*}

The workflow generates the data required to make Figure~4 of Abbott {\it et al.}~\cite{Abbott:2016blz}, however it does not generate the actual plot; a separate Python plotting script was used to create the histogram. As noted earlier, this script was not made public. Even though we had an internal version of the script, no PyInstaller bundle was created that captured the software stack used by that script. Running the plotting script against current versions of the libraries resulted in failures, so we needed to reproduce the original software stack. This was a considerable challenge and illustrates the importance of releasing containerized executables in addition to the source code for reproducibility in scientific analyses.

We obtained the version of PyCBC and LALsuite used by this code from notes made at the time of the original analysis (v1.3.4 and v6.36, respectively). We then determined the necessary and sufficient set of lower-level libraries required by these high-level libraries by examining the \texttt{setup.py} and \texttt{requirements.txt} in PyCBC v1.3.4. Using the PyCBC v1.3.4 install instructions, we create a Python virtual environment with the same version of \texttt{pip} and \texttt{virualenv} used in the original analysis. An iterative process of running the LALSuite \texttt{configure} script was performed until all the required dependencies of LALSuite were installed. The specific versions of fourteen libraries (and their dependencies) were either installed using \texttt{pip} or compiled from source into the Python virtual environment. The iterative process of determining the required dependencies was complicated by the fact that \texttt{pip} caches previous software builds and so the install process is not necessarily idempotent. 

After these libraries were installed, LALSuite and PyCBC were installed and the Python plotting script was executed. Our data release includes a script \texttt{make\_pycbc\_hist.sh}\footnote{\url{https://doi.org/10.5281/zenodo.4085984}} that automates the installation and execution of the plotting code. Our reproduction of the LIGO result is shown in Figure~\ref{f:gw150914-fig4b}, which includes the original LIGO/Virgo result for comparison.

We find that we were able to reproduce the search result. However, there are some small but noticeable differences in the search background (continuous black line) and the lower bound on the significance that the workflow reports for GW150914; We find the significance is greater than $5\sigma$, rather than greater than $5.1\sigma$ (original plot). We attribute both of these differences to changes in the input data used by the workflow. 
The usage instructions for the GWOSC data state that the LIGO strain data in the public data set are based on the \texttt{C02} calibration of the LIGO detectors, whereas the original PyCBC configuration files state that \texttt{C01} data was used for the analysis of Abbott {\it et al.}~\cite{Abbott:2016blz}. We hypothesize that the GWOSC \texttt{C02}-based data contains slightly less analysis time than the \texttt{C01} data originally used. This would result in a lower bound for the significance of the event and produce slight changes in the search background. However, we are not able to verify this as we were unable to obtain access to the proprietary \texttt{C01} data.

\section{CONCLUSIONS}
\label{sec:conclusion}

We have described the process and challenges encountered in reproducing the measured statistical significance of GW150914. Our script is configured to execute compute-intensive jobs on the OSG. To allow scientists to run on other resources, our data release provides instructions for running all jobs on local resources. The memory and runtime profiling of the workflow tasks provided in this paper will enable appropriate resource selection. 

Our execution of the workflow used HTCondor as the job scheduler; this scheduler is also used by the LIGO Data Grid. Although one could modify our workflow generation script to use alternative job schedulers, we recommend against this because of the wide variance in the memory requirements of the jobs and need for a relatively homogeneous environment. We rely on HTCondor for job resubmission in case of failure and its mechanisms of custom-created HTCondor classads to increase memory requested for a job in case of failures. When using a scheduler like SLURM, we recommend that one uses the upper memory bounds we provide in this paper. A better practice would be to overlay HTCondor on the native scheduler using resource provisioning techniques such as HTCondor glideins. We also recommend the use of CVMFS to access GWOSC data. Although Pegasus can be configured to transfer the data at runtime, e.g. from the submit host, where the workflow system is located (USC/ISI in our case) or via http from the GWOSC web site, this requires movement of tens of thousands of input data files.  It is more efficient to rely on the CVMFS storage and caching mechanism and to configure Pegasus to create symbolic links to the CVMFS locations, rather than performing true copies. 

We have demonstrated that, although LIGO did not provide complete instructions for reproducing the GW150914 result, sufficient information exists either in the public domain or recorded as notes describing the original analysis to reproduce the PyCBC GW150914 workflow. Although we made substantial progress in reproducing the PyCBC result shown in Figure~4 of Ref.~\cite{Abbott:2016blz}, we were unable to reproduce it exactly as we did not have access to the original input data and metadata. The LIGO data needs to be calibrated based on the understanding of the characteristics of the instrument and its state. These calibrations may change over time as the knowledge about the detectors improves. Data providers often want to publish ``best quality'' data and not provide earlier outdated versions. This is the case with the data from LIGO's first observing run where only the final calibrated data is public. If the original data used in the GW150914 discovery paper are made public, it is straightforward to modify our workflow generation script to use these data. As part of this work, we have released scripts that allows other scientists to reproduce the LIGO analysis using publicly available data using their own compute resources or the OSG using the latest stable versions of Pegasus and HTCondor.

Our results show that, in principle, it is possible to release instructions and code that allow other scientists to reproduce a major scientific result. We encourage scientists who wish to do so to ensure that instructions include: access to the original data and codes used; documentation of software and configuration file versions; containerized executables that capture the complete software stack used in the original analysis; long-term archival of code and data products used; and documentation about the computational resources needed to execute the analysis.
Understanding how reproducibility is incorporated in astrophysics workflows in general and scientific workflows in particular through the sharing of practices in reproducible scientific software will help enable open science across disciplines. Codes, data, and workflows generated by this and similar efforts can ultimately enable researchers and students at various levels of education to regenerate the same findings, learn about the scientific methods, and engage in new science, technology, engineering, and mathematics (STEM) research.

\section{ACKNOWLEDGMENT}

The scripts and supporting codes used to run the PyCBC workflow generation code using the GWOSC data and to make the result plot described in this article are available from GitHub at 
\url{https://github.com/gwastro/gw150914-fig4b}. The specific version used in this work was \url{https://doi.org/10.5281/zenodo.4085984}. The PyCBC v1.3.2 PyInstaller bundles used by the workflow and the HDF5 file created by the PyCBC workflow are available from \url{http://dx.doi.org/10.21227/c634- qh33}.

We would like to thank Mats Rynge for providing input on configuring the pipeline to run on OSG, Alexander Nitz and Maria Alessandra Papa for providing the script used to make the PyCBC result histogram, and Stuart Anderson for helpful discussions.
This work was supported by the U.S. National Science Foundation under Grants  OAC-1823378, OAC-1823405, OAC-1841399, and OAC-1823385. Pegasus is supported by the U.S. National Science Foundation under Grant OAC-1664162. The Open Science Grid is supported in part by the U.S. National Science Foundation under Grant PHY-1148698, and the U.S. Department of Energy’s Office of Science.  This research has made use of data, software and/or web tools obtained from the Gravitational Wave Open Science Center, a service of LIGO Laboratory, the LIGO Scientific Collaboration and the Virgo Collaboration. LIGO is funded by the U.S. National Science Foundation. Virgo is funded, through the European Gravitational Observatory (EGO), by the French Centre National de Recherche Scientifique (CNRS), the Italian Istituto Nazionale della Fisica Nucleare (INFN) and the Dutch Nikhef, with contributions by institutions from Belgium, Germany, Greece, Hungary, Ireland, Japan, Monaco, Poland, Portugal, Spain. 

%\balance
%\bibliographystyle{IEEEtran}
%\balance
%\bibliography{references}


\begin{IEEEbiography}{Duncan A. Brown}{\,} is the Charles Brightman Professor of Physics at Syracuse University, Syracuse, NY, USA. Dr.~Brown received a Ph.D. degree in physics from the University of Wisconsin-Milwaukee in 2004. He was a member of the LIGO Scientific Collaboration from 1999 to 2018 and is a fellow of the American Physical Society. He is a co-author of the paper ``Data  access  for LIGO on the OSG,'' which won Best Software and Data Paper at PEARC17. His research is in gravitational-wave astronomy and astrophysics, and the use of large-scale scientific workflows. Contact him at dabrown@syr.edu.
\end{IEEEbiography}

\begin{IEEEbiography}{Karan Vahi}{\,}is a Senior Computer Scientist at USC Information Sciences Institute, Marina Del Ray, CA, USA. Karan received a M.S in Computer Science in 2003 from University of Southern California. He is a co-author of the paper ``Integrity Protection for Scientific Workflow Data: Motivation and Initial Experiences'' which won Best Paper in the Advanced Research Computing Software and Applications Track and also the ''The Phil Andrews Most Transformative Contribution Award'' at PEARC19. His research interests include scientific workflows and distributed computing systems. Contact him at vahi@isi.edu.
\end{IEEEbiography}

\begin{IEEEbiography}{Michela Taufer}{\,} holds the Jack Dongarra Professorship in High Performance Computing within the Department of Electrical Engineering and Computer Science at the University of Tennessee, Knoxville. Dr. Taufer received her Ph.D. in computer science from the Swiss Federal Institute of Technology (ETH) in 2002. She is an ACM Distinguished Scientist and an IEEE Senior Member. Her interdisciplinary research is at the intersection of computational sciences, high permanence computing and data analytics. Contact her at taufer@acm.org.
\end{IEEEbiography}

\begin{IEEEbiography}{Von Welch}{\,} is the acting associate vice president for Information security, executive director for the OmniSOC, executive director for cybersecurity innovation at Indiana University, and the director of IU’s Center for Applied Cybersecurity Research (CACR). He specializes in cybersecurity for distributed systems, particularly scientific collaborations and federated identity.  Contact him at vwelch@iu.edu.
\end{IEEEbiography}

\begin{IEEEbiography}{Ewa Deelman}{\,} received her PhD in Computer Science from the Rensselaer Polytechnic Institute. She is a Research Director at USC/ISI and a Research Professor at the USC Computer Science Department  Her research explores the interplay between automation and the management of scientific workflows that include resource provisioning and data management. Her group has lead the design and development of the Pegasus Workflow Management software (http://pegasus.isi.edu) and conducts research in job scheduling and resource provisioning in distributed systems, workflow performance modeling, provenance capture, and the use of cloud platforms for science. Dr. Deelman is an AAAS and IEEE Fellow.  Contact her at deelman@isi.edu.
\end{IEEEbiography}

\end{document}